\documentclass[traditabstract,longauth]{aa}

\usepackage{graphicx}
\usepackage{txfonts}

\begin{document}

\title{{\it Herschel} ATLAS: The cosmic star formation history of quasar host 
galaxies\thanks{{\it Herschel} is an ESA space observatory with science
instruments provided by European-led Principal Investigator
consortia with important participation from NASA.}}

\author{S. Serjeant\inst{1}
\and F. Bertoldi\inst{29}
\and A.W. Blain\inst{24}
\and D.L. Clements\inst{2}
\and A. Cooray\inst{3}
\and L. Danese\inst{14}
\and J. Dunlop\inst{7}
\and L. Dunne\inst{4}
\and S. Eales\inst{8}
\and J. Falder\inst{5}
\and E. Hatziminaoglou\inst{6}
\and D.H. Hughes\inst{28}
\and E. Ibar\inst{7}
\and M.J. Jarvis\inst{2}
\and A. Lawrence\inst{7}
\and M.G. Lee\inst{25}
\and M. Micha{\l}owski\inst{21}
\and M. Negrello\inst{1}
\and A. Omont\inst{23}
\and M. Page\inst{26}
\and C. Pearson\inst{22,27,1}
\and P.P. van der Werf\inst{30}
\and G. White\inst{1,22}
\and A. Amblard\inst{4}
\and R. Auld\inst{8}
\and M. Baes\inst{9}
\and D.G. Bonfield\inst{2}
\and D. Burgarella\inst{10}
\and S. Buttiglione\inst{20}
\and A. Cava\inst{12}
\and A. Dariush\inst{8}
\and G. de Zotti\inst{20,14}
\and S. Dye\inst{8}
\and D. Frayer\inst{13}
\and J. Fritz\inst{9}
\and J. Gonzalez-Nuevo\inst{14}
\and D. Herranz\inst{15}
\and R.J. Ivison\inst{7}
\and G. Lagache\inst{16}
\and L. Leeuw\inst{17}
\and M. Lopez-Caniego\inst{15}
\and S. Maddox\inst{5}
\and E. Pascale\inst{8}
\and M. Pohlen\inst{8}
\and E. Rigby\inst{5}
\and G. Rodighiero\inst{11}
\and S. Samui\inst{14}
\and B. Sibthorpe\inst{7}
\and D.J.B. Smith\inst{5}
\and P. Temi\inst{17}
\and M. Thompson\inst{2}
\and I. Valtchanov\inst{18}
\and A. Verma\inst{19}
}


\institute{
$^{1}$Dept. of Physics and Astronomy, The Open University, Milton Keynes, MK7 6AA, UK\\ 
$^{2}$Astrophysics Group, Imperial College, Blackett Laboratory, Prince Consort Road, London SW7 2AZ, UK\\ 
$^{3}$Dept. of Physics \& Astronomy, University of California, Irvine, CA 92697, USA\\ 
$^{4}$School of Physics and Astronomy, University of Nottingham, University Park, Nottingham NG7 2RD, UK\\ 
$^{5}$Centre for Astrophysics Research, Science and Technology Research Institute, University of Hertfordshire, Herts AL10 9AB, UK\\ 
$^{6}$European Southern Observatory, Karl-Schwarzschild-Str., 85748 Garching b. Muenchen, Germany\\ 
$^{7}$UK Astronomy Technology Center, Royal Observatory Edinburgh, Edinburgh, EH9 3HJ, UK\\ 
$^{8}$School of Physics and Astronomy, Cardiff University, The Parade, Cardiff, CF24 3AA, UK\\ 
$^{9}$Sterrenkundig Observatorium, Universiteit Gent, Krijgslaan 281 S9, B-9000 Gent, Belgium\\ 
$^{10}$Laboratoire d'Astrophysique de Marseille, UMR6110 CNRS, 38 rue F. Joliot-Curie, F-13388 Marseille France\\ 
$^{11}$University of Padova, Department of Astronomy, Vicolo Osservatorio 3, I-35122 Padova, Italy\\ 
$^{12}$Instituto de Astrof\'{i}sica de Canarias, C/V\'{i}a L\'{a}ctea s/n, E-38200 La Laguna, Spain\\ 
$^{13}$National Radio Astronomy Observatory,  PO Box 2, Green Bank, WV  24944, USA\\ 
$^{14}$Scuola Internazionale Superiore di Studi Avanzati, via Beirut 2-4, 34151 Trieste, Italy\\ 
$^{15}$Instituto de F\'isica de Cantabria (CSIC-UC), Santander, 39005, Spain\\ 
$^{16}$Institut d'Astrophysique Spatiale (IAS), Bâtiment 121, F-91405 Orsay, France; and Université Paris-Sud 11 and CNRS (UMR 8617), France\\ 
$^{17}$Astrophysics Branch, NASA Ames Research Center, Mail Stop 245-6, Moffett Field, CA 94035, USA\\ 
$^{18}${\it Herschel} Science Centre, ESAC, ESA, PO Box 78, Villanueva de la Ca\~nada, 28691 Madrid, Spain\\ 
$^{19}$Oxford Astrophysics, Denys Wilkinson Building, University of Oxford, Keble Road, Oxford, OX1 3RH\\ 
$^{20}$INAF - Osservatorio Astronomico di Padova, Vicolo Osservatorio 5, I-35122 Padova, Italy\\ 
$^{21}$Scottish Universities Physics Alliance, Institute for Astronomy, University of Edinburgh, Royal
Observatory, Edinburgh, EH9 3HJ, UK\\ 
$^{22}$Space Science \& Technology Dept., CCLRC Rutherford Appleton Laboratory, Oxfordshire, OX11 0QX, UK\\ 
$^{23}$Institut d'Astrophysique de Paris, 98 bis boulevard Arago, 75014 Paris, France\\ 
$^{24}$Caltech, MS247-19, 1200 East California Blvd., Pasadena, CA 91125, USA\\ 
$^{25}$Astronomy Program, Department of Physics and Astronomy, Seoul National University, 
Shillim Dong, Kwan-ak Gu 151-742, Seoul, Korea\\ 
$^{26}$University College London, Department of Space \& Climate Physics, 
Mullard Space Science Laboratory, Holmbury St. Mary, Dorking, Surrey RH5 6NT, UK\\ 
$^{27}$Institute for Space Imaging Science, University of Lethbridge, Lethbridge, 
Alberta T1K 3M4, Canada\\
$^{28}$Instituto Nacional de Astrof\'{i}sica, \'{O}ptica y Electr\'{o}nica (INAOE),
Luis Enrique Erro No.1, Tonantzintla, Puebla, C.P. 72840, Mexico\\
$^{29}$Argelander Institut f\"{u}r Astronomie, Universit\"{a}t Bonn, Auf dem H\"{u}gel 71, Room 1.22, 53121 Bonn\\
$^{30}$Leiden Observatory, Leiden University, P.O. Box 9513, NL - 2300 RA Leiden, The Netherlands\\
}

\titlerunning{H-ATLAS: cosmic SFH of QSO host galaxies} 
\authorrunning{S. Serjeant et al.}

\abstract {We present a derivation of the star formation rate per comoving volume 
of quasar host galaxies, derived from stacking analyses of far-infrared to mm-wave photometry
of quasars with redshifts $0<z<6$ and absolute $I$-band magnitudes $-22>I_{\rm AB}>-32$ 
We use the science demonstration observations of the first $\sim16$\,deg$^2$ from
the {\it Herschel} Astrophysical Terahertz Large Area Survey (H-ATLAS) in which there are $240$ quasars
from the Sloan Digital Sky Survey (SDSS) and a further $171$ from the 2dF-SDSS LRG and QSO (2SLAQ) 
survey. We supplement this data with a compilation of data from IRAS, ISO, {\it Spitzer}, SCUBA and MAMBO. 
H-ATLAS alone statistically detects the quasars in its survey area at $>5\sigma$ at $250,350$ and 
$500\,\mu$m. From the compilation as a whole we find striking evidence of downsizing in quasar 
host galaxy formation: 
low-luminosity quasars with absolute magnitudes in the range $-22>I_{\rm AB}>-24$ have a comoving
star formation rate (derived from $100\,\mu$m rest-frame luminosities) 
peaking between redshifts of $1$ and $2$, while high-luminosity quasars with
$I_{\rm AB}<-26$ have a maximum contribution to the star formation density at $z\sim 3$. 
The volume-averaged star formation rate of $-22>I_{\rm AB}>-24$ quasars evolves as $(1+z)^{2.3\pm0.7}$
at $z<2$, but the evolution at higher luminosities is much faster reaching $(1+z)^{10\pm 1}$ at 
$-26>I_{\rm AB}>-28$. We tentatively interpret this as a combination of a declining major merger
rate with time and gas consumption reducing fuel for both black hole accretion and star formation.
}


\maketitle

\section{Introduction}

The cosmic star formation history (e.g. Madau et al. 1996 and
others) was quickly realised to bear a striking apparent similarity to
the evolving luminosity density of quasars (QSOs) 
at most redshifts (e.g. Boyle \& Terlevich
1998, Franceschini et al. 1999), suggesting a link between the
physical drivers of black hole growth and stellar mass assembly.
Other data also indirectly suggested links.
Mid-infrared spectra (e.g. Genzel et al. 1998, Spoon et al. 2007)
and radiative transfer modelling (e.g. Farrah et al. 2002) imply
higher luminosity starbursts have increasingly large bolometric
fractions from active galactic nuclei (AGN). The tight $K$-band Hubble
diagram of far-infrared-selected hyperluminous starbursts also bears a
striking similarity to the $K-z$ relation of radiogalaxies
(e.g. Serjeant et al. 2003, though see 
Smail et al. 2004 for their submm-selected counterparts). 

%

A close relationship between black hole growth and stellar mass
assembly is also demanded by the observed close correlations betwen
supermassive black hole masses and spheroid properties (e.g. Magorrian
et al. 1998, Merritt \& Ferrarese 2001), which exist despite the
enormous disparities of spatial scales and masses.  Such correlations
are predicted by models with radiative and/or kinetic energy outputs 
from the AGN regulating the star formation in their host galaxies
(e.g. Granato et al. 2006), but these feedback models have many adjustable
parameters. Feedback is arguably the principal uncertainty in
semi-analytic models of galaxy evolution. One of the few observational
approaches available to constrain feedback models is to measure the 
star formation rates in QSO host galaxies, but so far very few
QSOs have direct far-infrared, submm or mm-wave detections from
which star formation rates could be inferred. 


The {\it Herschel} Astrophysical Terahertz Large Area Survey (H-ATLAS, Eales
et al. 2010) is the largest open time key project on the {\it Herschel}
Space Observatory (Pilbratt et al. 2010). The survey aims to map
$550\,$deg$^2$ at five wavelengths from $110-500\,\mu$m to $5\sigma$
depths in the range $32-50$\,mJy at $\ge250\,\mu$m.  Among the key science goals is a 
constraint on the star formation rates of many thousands of QSOs
through far-infrared and submm photometry. In preparation, Serjeant \&
Hatziminaoglou (2009) used a compilation of available far-infrared to
mm-wave photometry of QSOs from IRAS, ISO, {\it Spitzer}, SCUBA and
MAMBO, to predict the numbers of QSOs directly detectable by
H-ATLAS. Several hundred QSO direct detections are expected in
H-ATLAS and the first H-ATLAS detections are described in
e.g. Gonzalez-Nuevo et al. (2010).  Serjeant \& Hatziminaoglou (2009)
also assumed an M82 spectral energy distribution (SED), used
observations close to $100(1+z)\,\mu$m where possible, and found the
stacked $100\,\mu$m luminosities of QSOs typically scaling roughly as the
square root of the optical luminosities with a redshift-dependent
normalisation, supporting the idea of coupled black hole mass and stellar
mass assemblies. Almost identical results were obtained with an Arp\,220
SED. This SED insensitivity can be readily understood:
at a fixed bolometric luminosity, the SWIRE template SEDs (Polletta et
al. 2007) for the starbursts M82, Arp\,220, NGC\,6090,
IRAS\,20551$-$4250 and IRAS\,22491$-$1808 have $100\,\mu$m
monochromatic luminosities all within a factor of two (note that we
are only concerned with the {\it starburst} bolometric contribution). 


In this paper we extend this analysis to the first data from 
H-ATLAS. We interpret the $100\,\mu$m luminosities as star
formation, since AGN dust tori SEDs are expected to peak in the mid-infrared, 
(e.g. Efstathiou \& Rowan-Robinson 1995, though see the discussion in 
Netzer et al. 2007). We
estimate the mean star formation rate in bins of redshift and absolute
$I_{\rm AB}$ magnitude, then use the QSO luminosity function to 
make the first constraints on the star formation rate per comoving volume 
of QSOs from $0<z<6$ and a factor of $10^4$ in optical luminosity. We assume
density parameters $\Omega_{\rm M}=0.3$ and $\Omega_\Lambda=0.7$ and a Hubble constant
of $H_0=70$\,km\,s$^{-1}$\,Mpc$^{-1}$. $I$-band QSO magnitudes are quoted in 
the AB system with $K$-corrections assuming 
${\rm d}\ln S_\nu /{\rm d}\ln \nu=-0.5$ and no internal dust extinction correction. 

\begin{figure}[!ht]
\vspace*{-1.2cm}
\begin{center}
   \resizebox{0.9\hsize}{!}{
     \includegraphics*[100,300][500,700]{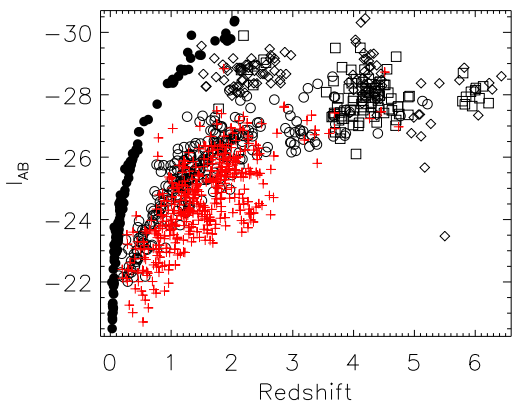}
   }
\end{center}
\vspace*{-2.2cm}
\caption{
The distribution of QSOs considered in this paper in the optical luminosity--redshift plane. 
Red crosses mark QSOs in the {\it Herschel} ATLAS survey. 
Open circles are SDSS QSOs with 
SWIRE coverage, while filled 
circles are Palomar-Green QSOs with IRAS and B-band data. Diamonds represent QSOs observed 
at $850\,\mu$m while open squares represent QSOs observed at $1200\,\mu$m. Adding
H-ATLAS QSOs extends the 
range of absolute magnitudes at redshifts $z<3$. 
}\label{fig:luminosity_redshift}
\vspace*{-0.25cm}
\end{figure}

\section{Data acquisition and analysis}\label{sec:data_acquisition}

\subsection{{\it Herschel} ATLAS}
We use only data from the SPIRE instrument (Griffin et al. 2010) in
this paper. For more details of the data analysis see 
Pascale et al. 2010 in prep.; we summarise the main points here.  The SPIRE images were
registered to a common reference frame using stacking analyses of
Sloan Digital Sky Survey (SDSS) galaxies. Neptune was used for 
flux calibration, requiring 
multiplicative calibration changes of 
$1.02$, $1.05$ and
$0.94$ at $250$, $350$ and $500\,\mu$m respectively relative to the previous
Ceres flux calibration (Griffin et al. 2010, Swinyard et al. 2010). Jumps in
thermometer timelines were identified and corrected using the mean
levels on either side of each jump.  Thermal drift modelling was
achieved through low-pass filtering. Bolometer timelines were filtered
using high-pass 4\,mHz filtering to remove correlated $1/f$ noise.
Maps were made using the naive map maker in HIPE and filtered
optimally for point sources.

\subsection{Quasar compilations}
The survey areas of H-ATLAS were chosen for their supplementary
multi-wavelength coverage. The H-ATLAS science demonstration field is covered by
the SDSS with $240$ QSOs in its spectroscopic QSO catalogue
in the field. There are also $171$ QSOs from 
the 2dF-SDSS LRG and QSO (2SLAQ) survey. 

We supplement our H-ATLAS QSOs with the compilation from Serjeant \& Hatziminaoglou (2009), 
including the authors' IRAS ADDSCAN photometry of Palomar-Green QSOs and {\it Spitzer} photometry
of QSOs from the {\it Spitzer} SWIRE legacy survey (Lonsdale et al. 2003, 2004), together with
QSOs observed at $850\,\mu$m or $1200\,\mu$m 
(Omont et al. 1996, Omont et al. 
2001, Carilli et al. 2001, Isaak et al. 2002, Omont et al. 
2003, Priddey et al. 2003a, 2003b, Robson et al. 2004, Wang 
et al. 2007). Figure \ref{fig:luminosity_redshift} shows the distribution of our QSO
compilation in the optical luminosity--redshift plane. Note that by combining several 
flux-limited samples we mitigate the effects of Malmquist bias (Teerikorpi 1984), 
i.e. that in a single flux-limited
sample there is a strong degeneracy between luminosity and redshift, making it 
impossible to decouple evolution effects from luminosity dependence. For the time
being we use the working assumption that any subtle biases in e.g. the Palomar-Green
sample are not strongly correlated with far-infrared luminosity; in time the 
full H-ATLAS QSO sample will allow us to make use of a more uniform QSO
selection. 


\section{Methods}\label{sec:methods}
We adopt the methodology of Serjeant et al. (2004, 2008, 2009) for comparing QSO fluxes 
with the flux distributions of the H-ATLAS maps as a whole. We use unweighted means to estimate 
H-ATLAS QSO fluxes and use the Kolmogorov-Smirnoff statistic to compare flux distributions. 
The advantage of this statistic is that it contains its own control test, i.e. there is no need for
performing stacks on randomised QSO positions. 

While most QSOs can only be detected in the far-infrared and submm through stacking analyses, 
there are a few QSOs with direct detections. How should one combine these? For example, if one
has a single non-detection of $0\pm4$\,mJy and single detection of 
$100\pm4$\,mJy, what can one say about the average flux $\mu$ of this
population? Clearly the answer is not $\mu=50\pm2$\,mJy. We have adopted the methodology of Serjeant and
Hatziminaoglou (2009) and treated flux measurements of individual QSOs as attempts to determine
the population mean. The dispersion in the population is an error term on this measurement, which one
would add in quadrature to the noise on any individual measurement. We determine the population dispersion
from our data and made a simultaneous maximum-likelihood fit to the mean $\mu$ and dispersion $\sigma$. There is
no covariance between these parameters because of the independence of signal and noise. Where there
are fewer than $3$ objects being considered we set $\sigma=0.84\mu$. More details can be found in
Serjeant \& Hatziminaoglou (2009). 

We used this method to combine direct detections with non-detections. We also combine
the Serjeant \& Hatziminaoglou (2009) sample with our H-ATLAS data. We use the point-source-convolved
H-ATLAS maps (Pascale et al. 2010). Measurements taken off these maps report the point source flux plus 
a background flux contribution from other galaxies. 
To set the latter to zero we convolve our maps with a further kernel $K$
with a zero total, i.e. $\int_0^\infty K(r)2\pi r{\rm d}r=0$.
$K$ was set to a constant negative value at radii $r$ of $2.5$ to $6$ times the point spread function
FWHM, with a unit $\delta$ function in its central pixel, and zero everywhere else. After convolution with $K$, 
every galaxy makes an exactly zero net contribution to the map. 
If there is a non-zero angular cross-correlation function from galaxies with a physical association with
the QSO, then they may still contribute to the far-infrared and submm fluxes. 
One approach is to use constraints on the
angular correlation function of these galaxies to place a bound on this contribution (e.g. Serjeant et al. 2008). 
However, in this case we are studying the assembly of the stellar mass and the companion galaxies may in
time be accreted by the QSO host galaxy. We have therefore chosen to associate all the star formation inferred 
from the flux in the far-infrared beam with the QSO host galaxy. 

\section{Results}
\subsection{Stacking analysis results}

We compared the flux measurements at the positions of our H-ATLAS QSOs with the distribution
of flux measurements throughout the H-ATLAS maps. 
The Kolmogorov-Smirnoff statistic rejects the null hypothesis that
these are drawn from the same distribution at significance levels of 
$3\times10^{-33}$, $7\times10^{-20}$ and $7\times10^{-9}$ at $250$, $350$ and $500\,\mu$m respectively, 
equivalent to $12.0\sigma$, $9.1\sigma$ and $5.7\sigma$. This is not due simply to the
presence of bright sources, since removing QSOs with fluxes $>100$\,mJy and comparing with
regions of the map with flux measurements $<100\,$mJy still yields significance levels of 
$11.9\sigma$, $9.0\sigma$ and $5.6\sigma$. 
The average fluxes of the QSOs in the H-ATLAS science demonstration field are 
$S_{250\,\mu{\rm m}}=9.41\pm 0.88$\,mJy ($11\sigma$ detection), 
$S_{350\,\mu{\rm m}}=7.68 \pm  0.87$\,mJy ($8.9\sigma$ detection) and 
$S_{500\,\mu{\rm m}}=5.14 \pm  0.92$\,mJy ($5.6\sigma$ detection). 


When calculating $100\,\mu$m rest-frame luminosities for H-ATLAS QSOs, we use the
closest SPIRE filter to $100(1+z)\,\mu$m. 
Table \ref{tab:qso_photometry} lists our estimates of the mean $100\,\mu$m rest-frame 
luminosities of the {\it whole} QSO compilation 
in redshift and optical luminosity bins, using the methodology 
of Sect. \ref{sec:methods}. The results are well-fit ($\chi^2_\nu=0.93$) by the expression 
$\log_{10}(\nu L_\nu(100\,\mu{\rm m})/10^{12}L_\odot)=\alpha(z)[I_{\rm AB}+\beta(z)]$, where 
$\alpha(z)=(0.0371\pm0.0048)\times\min(z,4)-0.235\pm0.018$, 
$\beta(z)=(-1.19\pm0.30)\times\min(z,4)+27.42\pm0.37$, and $\min(z,4)=z$ at $z<4$ and $4$ otherwise. 
The slope of the luminosity-luminosity correlation is shallower at high redshift, 
as found by Serjeant \& Hatziminaoglou 2009 (see also Mullaney et al. 2010). 

QSO number densities $\Phi(I_{\rm AB},z)$ at these absolute magnitudes and
redshifts are already well-determined. 
We adopted the luminosity functions of Croom et al. 2004
at $z<2.1$ and the pure density evolution model of Meiksin 2005 at $z>3$, with 
an optical spectral index of $-0.5$ to transform to 
$I$-magnitudes. Between $z=2.1$ and $z=3$ we interpolate in $\log\Phi$
between these models at fixed optical luminosities. We assume 
the far-infrared is dominated by giant molecular clouds so
$\nu L_\nu(100\,\mu{\rm m})/10^{12}L_\odot\equiv 265 M_\odot$/year as appropriate for our
assumed SED and a Salpeter initial mass function from $0.1$ to $100\,M_\odot$
(Kennicutt 1998). 
Table \ref{tab:qso_photometry} also lists the comoving star formation densities 
inferred from $\nu L_\nu(100\,\mu{\rm m})\times \Phi(I_{\rm AB},z)$. 

\begin{figure*}[!ht]
\vspace*{-7cm}
\begin{center}
\hspace*{-2cm}
   \resizebox{1.5\hsize}{!}{
     \includegraphics*[-130,0][1000,1000]{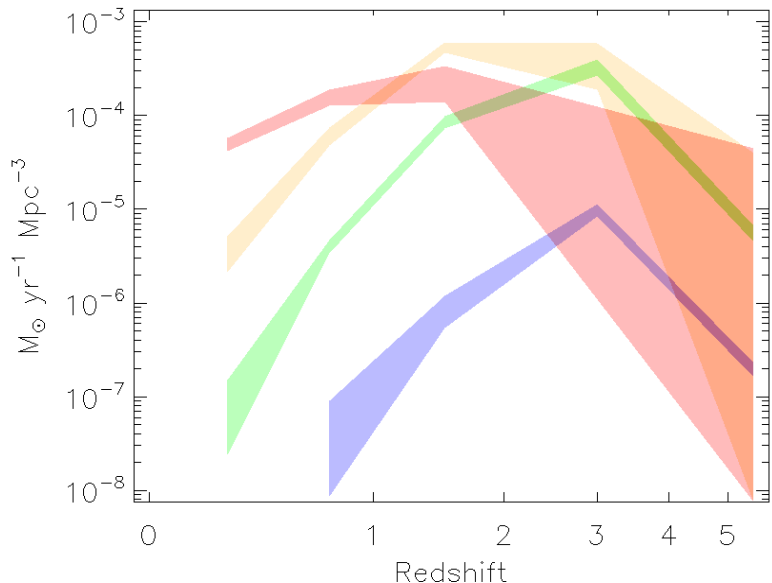}
   }
\end{center}
\vspace*{-9.5cm}
\caption{
Cosmic star formation history of QSO host galaxies inferred from $100\,\mu$m rest-frame
luminosities, for QSOs with $-22>I_{\rm AB}>-24$ (red), 
$-24>I_{\rm AB}>-26$ (orange), $-26>I_{\rm AB}>-28$ (green) and $I_{\rm AB}<-28$ (blue). 
The $2<z<4$, $-22>I_{\rm AB}>-24$ data point has too high a noise level to be
usefully constraining and has been omitted for clarity. 
For comparison, the $z=0$ total galaxy star formation rate is 
$(2.9\pm0.7)\times10^{-2}M_\odot$\,yr$^{-1}$\,Mpc$^{-3}$ 
(e.g. Serjeant et al. 2002). 
}\label{fig:madau}
\end{figure*}

Figure \ref{fig:madau} plots the data in table \ref{tab:qso_photometry}, interpolating between
the upper and lower bounds of the comoving star formation densities at the mid-points of
each redshift bin. The evolution in the comoving star formation density 
of $-22>I_{\rm AB}>-24$ QSOs from the $0.05<z<0.5$ bin to the $1<z<2$ bin 
scales as $(1+z)^{2.3\pm0.7}$,
but the evolution at higher luminosities is much faster: at $-24>I_{\rm AB}>-26$ the
variation is $(1+z)^{7.4\pm0.6}$, while at $-26>I_{\rm AB}>-28$ it reaches an
astonishing $(1+z)^{10\pm 1}$ over this redshift range (mostly but not entirely due
to the luminosity function). There are no QSOs at $I_{\rm AB}<-28$
at $z<0.5$ but between the $0.5<z<1$ and $1<z<2$ bins, the evolution of the volume-averaged
star formation in $I_{\rm AB}<-28$ QSOs scales as $(1+z)^{8\pm 3}$. An important caveat is
that we are only addressing the optically-defined QSO population, so we are necessarily
missing the type-2 QSOs and Compton-thick objects (see e.g. the discussion in Almaini, 
Lawrence \& Boyle 1999). 
A similar downsizing effect is seen in 
the QSO soft X-ray luminosity function (e.g. Hasinger et al. 2005).
The type-2 QSO fraction increases with redshift (e.g. Hasinger 2008)
which would increase our inferred evolution rates. 

\begin{table}
\begin{tabular}{lllllll}
$N$ & $z_{\rm min}$ & $z_{\rm max}$ & $I_{\rm min}$ & $I_{\rm max}$ & $\nu L_\nu(100\,\mu{\rm m})$ & SFD\\
    &              &             & (AB)         & (AB)         & $/10^{12}L_\odot$ & $/$($M_\odot$ yr$^{-1}$ Mpc$^{-3}$)\\
\hline 
 6  & $0.05$       & $0.5$       &$-22$         & $-19$        &  $0.126\pm 0.069$ & $(2.0\pm1.1)\times10^{-4}$\\ 
 72 & $0.05$       & $0.5$       &$-24$         & $-22$        &  $0.115\pm 0.019$ & $(4.94\pm0.79)\times10^{-5}$\\ 
 29 & $0.05$       & $0.5$       &$-26$         & $-24$        &  $0.31\pm  0.13 $ & $(3.6\pm1.5)\times10^{-6}$\\ 
 9  & $0.05$       & $0.5$       &$-28$         & $-26$        &  $0.52\pm  0.38$ & $(8.8\pm6.4)\times10^{-8}$\\ 
 9  & $0.5$        & $1$         &$-22$         & $-19$        &  $0.028\pm 0.065$ & $(0.5\pm1.2)\times10^{-4}$\\ 
 73 & $0.5$        & $1$         &$-24$         & $-22$        &  $0.173\pm 0.034 $ & $(1.58\pm0.31)\times10^{-4}$\\ 
 38 & $0.5$        & $1$         &$-26$         & $-24$        &  $0.400\pm 0.084 $ & $(6.1\pm1.3)\times10^{-5}$\\ 
 5  & $0.5$        & $1$         &$-28$         & $-26$        &  $1.54\pm  0.24$ & $(4.06\pm0.63)\times10^{-6}$\\ 
 2  & $0.5$        & $1$         &$-32$         & $-28$        &  $5.5\pm   4.5$ & $(4.9\pm4.0)\times10^{-8}$\\ 
 46 & $1$          & $2$         &$-24$         & $-22$        &  $0.218\pm  0.090 $ & $(2.4\pm1.0)\times10^{-4}$\\ 
 223& $1$          & $2$         &$-26$         & $-24$        &  $0.727\pm 0.092$ & $(5.38\pm0.69)\times10^{-4}$\\ 
 75 & $1$          & $2$         &$-28$         & $-26$        &  $1.81\pm  0.26 $ & $(8.6\pm1.2)\times10^{-5}$\\ 
 32 & $1$          & $2$         &$-32$         & $-28$        &  $5.1\pm   1.9 $ & $(8.6\pm3.2)\times10^{-7}$\\ 
 6  & $2$          & $4$         &$-24$         & $-22$        &  $1\pm17$ & $(0.4\pm8)\times10^{-3}$\\ 
 52 & $2$          & $4$         &$-26$         & $-24$        &  $1.05\pm  0.54 $ & $(3.9\pm2.0)\times10^{-4}$\\ 
 108& $2$          & $4$         &$-28$         & $-26$        &  $3.47\pm  0.68 $ & $(3.30\pm0.65)\times10^{-4}$\\ 
 66 & $2$          & $4$         &$-32$         & $-28$        &  $5.42\pm  0.76 $ & $(9.9\pm1.4)\times10^{-6}$\\ 
 1  & $4$          & $7$         &$-24$         & $-22$        &  $1.7\pm   1.9 $ & $(2.1\pm2.4\times10^{-5}$\\ 
 1  & $4$          & $7$         &$-26$         & $-24$        &  $2.0\pm   2.2 $ & $(1.9\pm2.1)\times10^{-5}$\\ 
 58 & $4$          & $7$         &$-28$         & $-26$        &  $2.32\pm  0.46 $ & $(5.6\pm1.1)\times10^{-6}$\\ 
 64 & $4$          & $7$         &$-32$         & $-28$        &  $4.25\pm  0.69$ & $(1.95\pm0.32)\times10^{-7}$\\ 
\end{tabular}
\caption{\label{tab:qso_photometry} 
Average stacked QSO $100\,\mu$m rest-frame luminosities for bins in redshift and $I$-band
absolute magnitude. An M82 SED has been assumed, though the results are only very weakly
dependent on the assumed SED. The first column gives the number of QSOs in the
bin in question. The final column gives the comoving volume-averaged star formation densities  
of QSO host galaxies in each bin. 
}
\end{table}


\section{Discussion and conclusions}
Could the co-evolution of the total cosmic star formation history and
total black hole accretion be explained by both being simultaneously
driven by major galaxy-galaxy mergers (e.g. di Matteo et al. 2005)?
Most star formation at $z<1$ does {\it not} seem to be 
triggered by major mergers (e.g. Bell et al. 2005), but QSO host
galaxies often have signs of significant disturbance. This suggests
a more complex picture. An alternative possibility proposed by Zheng
et al. (2009) is that star formation occurs mainly in disks, while
black hole growth occurs separately after major merger events. Any
apparent co-evolution in this model is present because the black hole
accretion depends on the potential well of the spheroid, which in turn
grows through the addition of disrupted disk stars.

Our observation of an unprecedented strong evolution in the star
formation rates of the brightest QSOs has no obvious analogue in
evolving galaxy populations. This is difficult to accommodate in
models with quasi-simultaneous starburst and AGN events sharing a
common trigger. Instead, it suggests star formation in bright QSO
hosts is 
not representative of stellar mass assembly in
general. We find QSO hosts comprise $\stackrel{<}{_\sim}1\%$ of
the total volume-averaged star formation density of the Universe at all
redshifts (e.g. Micha{\l}owski et al. 2010). 

What could drive the extremely strong 
downsizing 
evolution in bright QSOs? If major mergers
drive QSO activity, the declining major merger rate at a given halo
mass could account for the decreasing QSO number density.  However,
this is not enough to account for our observations, since we also find
that the relationship between optical luminosities (our proxy for
black hole accretion rate) and far-infrared luminosities (our proxy
for star formation rate) changes with redshift, in the sense that
bright QSOs at a {\it fixed} optical luminosity were more
far-infrared-luminous at high redshifts. One possibility is that
less gas is available at low redshift for star formation and 
black hole accretion, due to consumption in star formation or ejection
through feedback mechanisms. The strong decline in the comoving
star formation densities of bright QSOs could therefore be due to
a double dependence on the availability of gas combined with a declining 
major merger rate. 


\begin{acknowledgements} 
  We thank the anonymous referee for useful comments.
  This work was funded in part by STFC (grants PP/D002400/1 and
  ST/G002533/1).  Funding for the SDSS and SDSS-II has been provided
  by the Alfred P. Sloan Foundation, the Participating Institutions,
  the NSF, the U.S. Department of Energy, NASA, the Japanese
  Monbukagakusho, the Max Planck Society, and HEFCE.
 \end{acknowledgements}
 

%
%

%


\end{document}